\documentclass[reprint,12pt,onecolumn,notitlepage]{revtex4-1}

\usepackage{epsfig}
\usepackage{amsmath}   
\usepackage{amsfonts}
\usepackage{amssymb}
\usepackage{graphicx}  
\usepackage{verbatim}
\usepackage{color}   
\usepackage{subfigure}
\usepackage{hyperref}
\begin{document}

\title{Power laws in physics\\
	{\rm published in {\em Nature Reviews Physics} \bf{4} {\rm 501-503 (2022)}}}

\author{James P. Sethna}
\affiliation{LASSP, Physics Department, Cornell University, Ithaca, NY 14853-2501, United States}
\email{sethna@lassp.cornell.edu}
\date{\today}

\newcommand{\ddFlat}[2]{{{\mathrm{d} #1}/{\mathrm{d} #2}}}
\newcommand{\dd}[2]{{\frac{\mathrm{d} #1}{\mathrm{d} #2}}}
\newcommand{\curlyZ}{{\mathcal{Z}}}
\newcommand{\curlyP}{{\mathcal{P}}}
\newcommand{\curlyA}{{\mathcal{A}}}
\newcommand{\curlyR}{{\mathcal{R}}}
\newcommand{\curlyF}{{\mathcal{F}}}
\newcommand{\wt}{\widetilde}
\newcommand{\wb}[1]{\overline{{#1}}}

\begin{abstract}
Getting the most from power-law-type data can be challenging. James Sethna points out some of the pitfalls in studying power laws arising from emergent scale invariance, as well as important opportunities.
\end{abstract}

\maketitle

\thispagestyle{empty}

Power laws arise in many fields of knowledge -- from word usage in linguistics, to income distributions in economics. There is an
enormous literature observing and calculating power laws in nature. Publication of new, interesting results may involve data spanning one to two decades~\cite{AvnirBLM98}: we need good tools to show that the power laws are real and accurate. In short, power laws are easy
to fit, but challenging to measure and interpret
well~\cite{Newman05}. 
What are the particular challenges 
in studying power laws stemming from emergent scale 
invariance, a focus of much of statistical physics? And what opportunities exist
to extract more science from the data?

\section{Universal scaling functions}
Many systems show fractal structure and scale-invariant fluctuations
as they get large --- the rules describing their behavior look the same
up to rescaling as one observes larger and larger systems. Continuous phase
transitions (like the Curie point in ferromagnets), dynamical behavior of 
disordered systems (depinning transitions, crackling noise and
avalanches), the onset of chaos, earthquakes, fully developed turbulence,
and the behavior of the stock market all show clear symptoms of
emergent scale invariance, and all exhibit power laws in various measures
of their behavior. In many of these systems, these power
laws are explained using the renormalization group (RG)~\cite{SethnaDM01},
which coarse-grains a system
and then rescales (renorms) the parameters and observables to reach a fixed
point.
In some systems (turbulence, earthquakes) there is almost
  a consensus. In other systems (glasses~\cite{LiarteTSCCSnn},
  random matrix theory~\cite{AdamBSW02}) there are 
  universal critical exponents and universal scaling functions, with no
  known RG explanation.
The renormalization group predicts power laws relating various quantities,
which are universal --- shared between theory and experiment, and also shared
between strikingly different experimental systems in the same `universality
class'. If $Z$ depends on $X$, then
$Z \sim X^\beta$ for some usually non-trivial, probably transcendental,
universal critical exponent $\beta$.

The RG also predicts universal scaling functions for 
relations involving more than two parameter or observables.
If $Z$ depends on $X$ and $Y$, then 
\begin{equation}
\label{eq:UniversalScalingFunction}
Z(X,Y) \sim X^\beta \curlyZ(X/Y^\alpha)
\end{equation}
where $\alpha$ is also a universal number and $\curlyZ$ is a universal
function. The challenges and most fruitful opportunities for 
experimentalists and simulators in measuring these power laws almost 
invariably involve corrections and modifications of the power laws due 
to these powerful universal scaling functions. 

\section{Finite-size scaling and scaling collapses}
We start with finite-size scaling, describing the behavior in
a system confined to a cubic box of size $L$ (or in a material with 
grains of size $L$). Suppose our system exhibits
avalanches with sizes $S$ spanning a large range. Then the fraction of the
motion lying in avalanches with size between $S$ and $S+\mathrm{d}S$ is 
\begin{equation}
\label{eq:FiniteSizeScaling}
A(S,L) \sim S^{1-\tau} \curlyA(S/L^{d_{\rm f}}),
\end{equation}
where $d_{\rm f}$ is the fractal dimension of the avalanche, so an avalanche
spanning the system will have a typical size $S \sim L^{d_{\rm f}}$. 

\begin{figure}[ht]
\centering
\includegraphics[width=\linewidth]{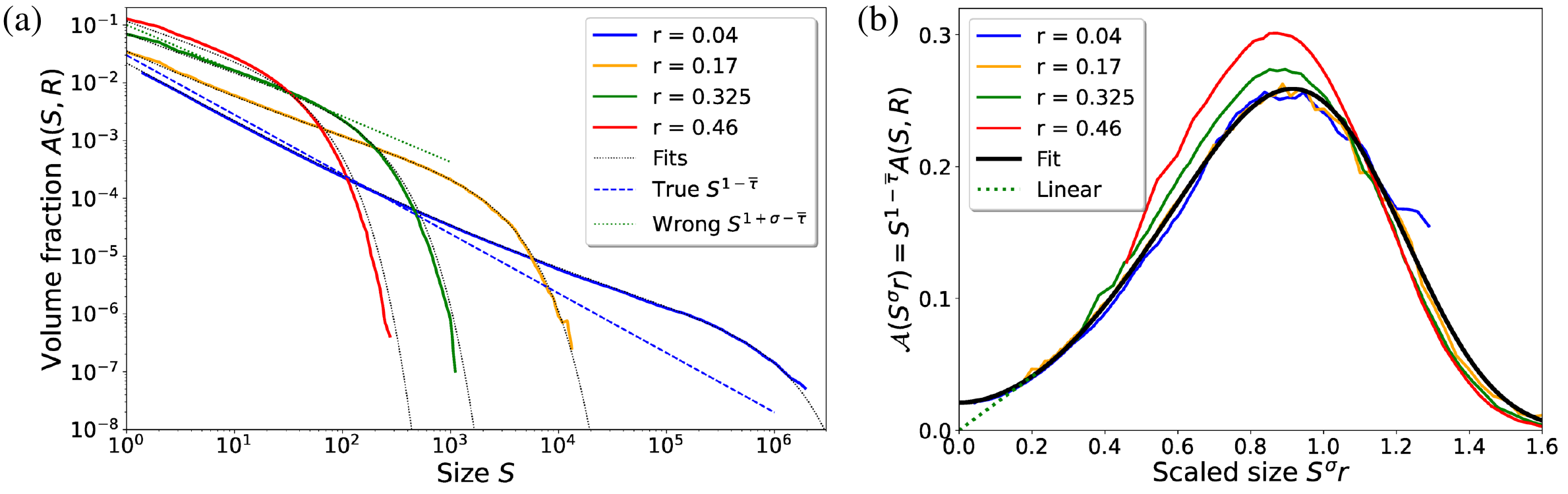}
\caption{
\label{fig:PowerLawsAndScalingCollapses}
{\bf Power laws and avalanche sizes in 
a random-field Ising model under an increasing field.}~\cite{PerkovicDS95}
(a)~Avalanche probability distribution $A(S,R)$ that a site is 
in an avalanche of size $S$, for disorder $R$. Data are plotted at different values of $r=(R-R_{\rm c})/R$, where $R_{\rm c}$ is the critical disorder. The true power-law exponent is $1-\wb{\tau}$; the apparent (wrong) power-law exponent is $1+\sigma-\wb{\tau}$. Note that one needs over four decades
of scaling to discover the correct power law. (b)~Scaling collapse
of the same data, together with a fit to the scaling function 
$\mathcal{A}(S^\sigma r)$. Corrections to scaling are responsible for the deviations far from $r=0$. Note that one needs simulations of a billion
spins to discover that the asymptote of $\mathcal{A}$ was non-zero: smaller simulations gave the wrong power law given by the dotted lines. Data reproduced
from~\cite{PerkovicDS95}.
}
\label{fig}
\end{figure}

It is natural
that avalanches larger than this will be strongly suppressed! So
$\curlyA$ will decrease quickly as its argument grows past one. Conversely,
if $\curlyA$ goes to a positive constant as its argument goes to zero,
then small enough avalanches will have the predicted universal power law
volume fraction $S^{1-\tau}$. But an experiment or simulation that measures
avalanches in a size region where $\curlyA$ is varying will often find a rather
good --- but incorrect --- power-law fit
(Fig.~\ref{fig:PowerLawsAndScalingCollapses}a).

A much better practice is to vary the system size (or the
grain size) and do a scaling collapse: plotting
$S^{\tau-1} F_L(S)$ against $S/L^{d_{\rm f}}$, and varying $\tau$ and $d_{\rm f}$ until all
the curves lie atop one another (Fig.~\ref{fig:PowerLawsAndScalingCollapses}b).

\section{Subdominant corrections and fitting functional forms}
Finite-size scaling produces corrections important when the behavior
reaches the system size. But what about corrections important for
small scales? Or when we are farther from the critical point? There are two types of `subdominant' corrections, namely 
singular corrections
to scaling and analytic corrections to scaling. For example, the liquid--gas
critical point has a free energy of the form
\begin{equation}
\label{eq:CorrectionsToScalingIsing}
F(T,P,u) \sim \wt{t}^{\beta + \beta \delta}
		\curlyF(\wt{h}/\wt{t}^{\beta \delta}, \wt{u}/t^{-\Delta})
\end{equation}
The variable $u$ is an irrelevant control variable: it is multiplied by 
zero at $t=0$, which becomes
less and less important as one approaches the critical temperature $T_{\rm c}$
and pressure $P_{\rm c}$. The functions
$\wt{t}(T,P,u)=a (T-T_{\rm c}) + b (P-P_{\rm c}) + c (T-T_{\rm c})^2 + \dots$, $\wt{h}(P,T,u)$,
and $\wt{u}(u,T,P)$ are analytic power series, that embody how temperature
and pressure map onto the `natural' RG parameters $t$,
$h$, and $u$ in the Ising universality class. 

By doing a Taylor expansion
in $u$, $T-T_{\rm c}$, and $P-P_{\rm c}$, one gets corrections that go as higher powers
of $T-T_{\rm c}$. In particular, the irrelevant variable $u$ causes a singular
correction to scaling which is $\wt{t}^\Delta$ times smaller than the 
dominant singularity. 

When we have scaling functions with more than one variable as in Eq.~\ref{eq:CorrectionsToScalingIsing}, scaling collapses
no longer are useful. A powerful, satisfying and numerically convenient
approach is to do a multiparameter fit to the data~\cite{ChenPSZD11,ChenZS15,HaydenRS19},
varying not only parameters like $\beta$, $\delta$, $u$ $T_{\rm c}$ and $P_{\rm c}$, $a$, $b$, $c$, and so on,  but also a parameterized functional form for the scaling function 
$\curlyF$. 

Fitting functional forms have three additional benefits. First, they
provide estimates not only of the universal critical exponents, but
also of the equally universal scaling functions. Second, they allow for estimates
of both statistical and systematic~\cite{ChenPSZD11} errors in the exponents
(which are often much larger than those of a straight power-law fit). 
Finally, these corrections, which are tiny near the critical point, become of
increasing importance for describing precursor fluctuations in the surrounding
phases. Indeed, here one imagines describing the (challenging) properties
of liquids far into the phase diagram using analytic and singular
corrections to the Ising critical point.

\section{Singular scaling function and dangerous irrelevant variables}
Being careful to measure properties on sizes large compared to microscopic
and small compared to the system, will one find the correct power laws?
Not if our scaling function is itself singular ---  going to zero or
infinity as its argument goes to zero. In a study by our group of the
random-field Ising model in 3D\cite{PerkovicDS95}, this almost happened
(Fig.~\ref{fig:PowerLawsAndScalingCollapses}).
We were measuring the fractional
coverage of avalanches $A(S,R) \sim S^{1-\wb{\tau}} \curlyA(S^\sigma r)$.
where $r=(R-R_{\rm c})$ is the distance to a critical disorder. We found
excellent scaling collapses, but $\curlyA$ seemed to go linearly to zero
as $S^\sigma r$ went to zero (dotted line in Fig.~\ref{fig:PowerLawsAndScalingCollapses}b) --- leaving us with an effective power law
$A(S,R) \sim S^{1-\wb{\tau}+\sigma}$ (dotted line in Fig.~\ref{fig:PowerLawsAndScalingCollapses}a) that disagreed with the `RG' exponent $1-\wb{\tau}$ extracted from the scaling collapse.
In the end, we used
(at the time) heroic billion-site simulations to discover that $\curlyA$ only
nearly vanishes --- it rises by a factor of ten from its small initial value.

Singular scaling functions also arise in 
the important case of dangerous irrelevant variables
--- quantities like $u$ in Eq.~\ref{eq:CorrectionsToScalingIsing}
that vanish under rescaling (are irrelevant), but for which the scaling function
for a physical properties diverges as it vanishes. This happens in 
some glassy systems, in which the freezing on long length scales is not
the usual competition between temperature and coupling between particles,
but instead a competition between random disorder and coupling. Temperature
acts only to hop over barriers, allowing the system to relax. Because
temperature is an irrelevant variable at the glass transition, the relaxation
time (and its scaling function) diverges as the system is cooled through
the transition.

\section{Crossover scaling, nonlinear RG flows, and all that}
There are many more fascinating implications and uses for universal scaling
functions, and associated warnings that fitting power laws can lead you
astray. Many systems exhibit crossovers, going smoothly from
one power law to another as the scales become large --- commonly arising
for quantum critical points observed at finite temperatures, but also
observed, for example, in magnetic avalanches~\cite{Sethna07Crossover} and
fracture and depinning transitions~\cite{ChenZS15}.
Other systems exhibit more complex 
scaling behavior, because their RG flows are intrinsically
nonlinear~\cite{HaydenRS19,RajuCHKLRS19}. This is remarkably common, for example,
at critical points in phase transitions, where all systems in 2D
and 4D have either logarithms, exponentials, or essential
singularities.

Thus the pitfalls of trusting a power-law fit should be viewed not as an
obstacle, but an opportunity. It is challenging, but intellectually and
scientifically fruitful, to extract the most from your data.

\bibliography{PowerLaws,SethnaRecs}

\section*{Acknowledgments}

\section*{Competing interests}
The author declares no competing interests.

\end{document}